# How emoji and word embedding helps to unveil emotional transitions during online messaging

Moeen Mostafavi
*Engineering Systems and Environment*
*University of Virginia*
Charlottesville, United States
moeen@virginia.edu

Michael D. Porter
*Engineering Systems and Environment*
*School of Data Science*
*University of Virginia*
Charlottesville, United States
mdp2u@virginia.edu

*Abstract*—During online chats, body-language and vocal characteristics are not part of the communication mechanism making it challenging to facilitate an accurate interpretation of feelings, emotions, and attitudes. The use of emojis to express emotional feeling is an alternative approach in these types of communication. In this project, we focus on modeling a customer's emotion in an online messaging session with a chatbot. We use Affect Control Theory (ACT) to predict emotional change during the interaction. To let the customer use emojis, we also extend the affective dictionaries used by ACT. For this purpose, we mapped Emoji2vec embedding to the affective space. Our framework can find emotional change during messaging and how a customer's reaction is changed accordingly.

*Index Terms*—Affect control theory, word embedding, emoji embedding, emotional state, interaction modeling

## I. Introduction

In personal communication, nonverbal cues transfer most of the meaning related to feelings and attitudes. Mehrabian [1] suggests that verbal communication only provides 7% of one's perceived attitude while 38% is based on vocal inflections and 55% from facial expressions. Despite the difficulties of accurately conveying emotions and attitudes, nonverbal communication, like messaging, has become a dominant mode of communication [2]. This study explores the components of interactions in the context of messaging and provides an approach to model an individual's emotion using a combination of words and emojis. Knowing the user's emotion helps machines such as chatbots have more naturalistic communication with humans.

Tracking the emotional states during a text communication requires a dynamic model. The model updates the states after each communication dialogue. We use Affect Control Theory (ACT) to model an individual's emotional transition in this process.

ACT is a formal theory of culture that attempts to explain social behavior. ACT predictions about interaction dynamics are dependent on initial cultural sentiments, which are indexed in the affective dictionaries. Using emoji and word embedding, we explain how we can extend affective dictionaries to include the most commonly used words and emojis in daily applications. Then we discuss how to improve the interpretation of emotions during online messaging interactions using extended affective dictionaries.

This project's contribution will be in two primary areas: (1) Developing an extended affective dictionary that includes emojis in their embedding. (2) Develop an algorithmic way to use emoji and word embedding for emotion change modeling during online messaging interactions.

## II. Background and Relevant prior work

### A. Affect Control Theory

According to ACT, all concepts in a given culture have shared emotional meanings that can be represented in three dimensions: Evaluation, Potency, and Activity (EPA). Evaluation contrasts "good" versus "bad", Potency contrasts "strong" versus "weak", and Activity contrasts "fast" versus "slow". EPA space, $[-4.3, 4.3]^3$, is a semantic word representation in a three-dimensional vector space. ACT assumes concepts can be indexed culturally using these three dimensions.

To model the consequences of an action or event, ACT uses an Actor-Behavior-Object (ABO) structure. To put it simply, an action involves an Actor directing Behavior to an Object. ACT assumes a given culture has shared EPA representations of Actor, Behavior, and Object known as *sentiments*. These sentiments are the baseline representation of the ABO characters in EPA space. However, an action can change the perception, or *impression* of an actor or object. That is, the perceived EPA values a member of the community assigns the actor or object may change as a result of the action or event.

Consider a simple example of a nurse (Actor) helping/shouting (Behavior) at a patient (Object). Within the culture, people have a sentiment representation of the nurse and the patient. Based on the action, EPA impressions may change from the default values. If a nurse is shouting at the patient, he is expected to be more active but not as good as other nurses. This impression for the nurse is different from the default sentiments. On the other hand, the patient is considered to be less powerful when the nurse shouts at him. In Fig. 1 the impression of these two events are shown.

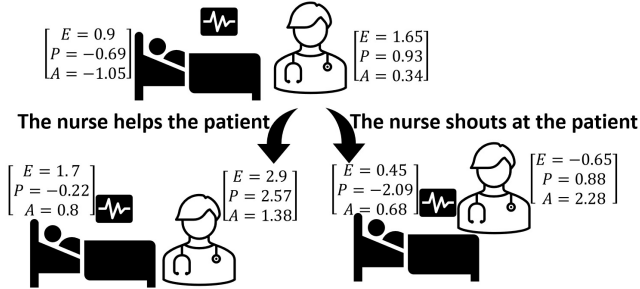

Fig. 1. Event impression - a nurse interacting with a patient. At the top of the this picture we can find the sentiments associated with a nurse and a patient character. Then we can find the impression for these two characters after the nurse helps/shouts at the patient in the bottom of the picture.

ACT uses impression change equations [3] similar to (1) to predict the transient meanings (impressions) that arise from ABO events.

$$C'_g = c + \sum_{D,h} \alpha_{Dh} D_h + \sum_{D,F,h,i} \beta_{D_h F_i} D_h F_i \quad (1)$$

where $c, \alpha, \beta$ are constant values, $\{C, D, F\}$, represent one of the ABO characters, and $\{g, h, i\} \in \{e, p, a\}$ represent the ABO character in one of EPA dimensions before the event, and $C'_g$ represent the resulting impression after the event.

$$A'_a = -0.26 + 0.41 A_e + 0.42 B_e + 0.03 O_e + 0.06 O_p \quad (2)$$
$$- 0.02 B_p - 0.10 B_a + 0.05 A_e B_e + 0.03 A_e O_p$$
$$- 0.05 B_e O_p - 0.05 B_p O_e + 0.12 B_e O_e$$
$$+ 0.03 A_e B_e O_e - 0.02 A_e B_e O_p$$

As an example, given the EPA sentiments for "shout at someone" is $[-1.08, 0.85, 1.89]$, (2) gives the activity impression of the nurse after shouting at the patient to be $-0.65$.

The nine equations similar to the equation (2) are known as impression change equations. Heise discussed finding impression change equations in [4].

The distance between the default sentiments and resulting impressions is termed *deflection*. Highly deflecting events create social and physiological distress [5]. Based on the ACT, the driving force in human interaction is minimizing the deflection [6]. In other words, individuals tend to feel distressed if the impression of their actions is different from their sentiments.

ACT explains how the cultural meanings of identities in an ABO event shape individuals' actions and experiences [7]. ACT provides a framework to model cultural identities and behaviors in groups [8]. Cultural meanings change over time, and different groups might have different cultural meanings. Affective meaning within the group influences the transmission of the information between the individuals. Cultural consistency biases play an essential role in information change [7]. Researchers used ACT to find response changes to social events due to language cultures in a diverse society [9].

Validated and applied in more than 100 studies [10], ACT has yet to be rigorously applied to online messaging. A major difficulty in incorporating ACT for modeling online messaging interactions is that the existing affective dictionaries are not sufficiently rich. Affective dictionaries come from surveys. Because surveys are time-consuming and expensive, most affective dictionaries focus on a culture's most commonly used words; often less than 3000 words. This covers only a tiny fraction of words commonly used in online messaging and ignores other communication forms like emojis. Due to the limited dictionary size, the application of ACT in modeling social interactions is severely restricted; only events that use words in the dictionaries can be analyzed.

### B. Interaction simulation

To better understand how ACT models the change of emotions and accordingly deflection changes, consider the following scenario of sequential interactions between a customer and a salesman in Table I.

TABLE I
INTERACTION OF A CUSTOMER AND A SALESMAN

| Index | Actor | Chat | Action |
|---|---|---|---|
| 1 | Customer | Hello | greet |
| 2 | Salesman | Hello, I am the Bot | welcome |
| 3 | Customer | Why my order status is not updated | grouse at |
| 4 | Salesman | What is your order number | question |
| 5 | Customer | order #8218 | answer |
| 6 | Salesman | Please wait, I check it | request sth. from |
| 7 | Customer | Take your time | agree with |
| 8 | Salesman | We shipped it and will be delivered in two weeks | answer |
| 9 | Customer | Not acceptable, it is too late | criticize |
| 10 | Salesman | Sorry, I can refund the shipping cost | gratify |
| 11 | Customer | I need full refund | argue with |
| 12 | Salesman | You get full refund in two business day | uplift |
| 13 | Customer | Sounds good | agree |
| 14 | Salesman | Thank you for contacting us | thank |
| 15 | Customer | Bye | leaves |

The INTERACT software [3] can simulate the deflection from social events by analyzing role behaviors, interpersonal emotions, identity and labeling processes, and trait attributions [3]. To simulate interactions, INTERACT combines the EPA sentiments and impression equations to produce the expected impression from ABO events. To simulate an interaction like this, we need to assign action/behavior tags to the chat dialogues. These tags must be selected from the software's affective dictionary. In Table I, the "Action" column includes potential tags for the "Chat" dialogues. Note that Action tags could be improved with an expanded affective dictionary. Simulating the above scenario using INTERACT produces the visualization in Table II. This shows how uses ACT to model the emotions of the interaction.

TABLE II
Facial change during the customer-salesman interaction [3]

INTERACT finds the deflection throughout interactions and plots it. It also finds emotions, attributes, and labels that minimize the deflection for the Actor and Object in this interaction. Fig. 2 includes a screenshot of its simulation I.

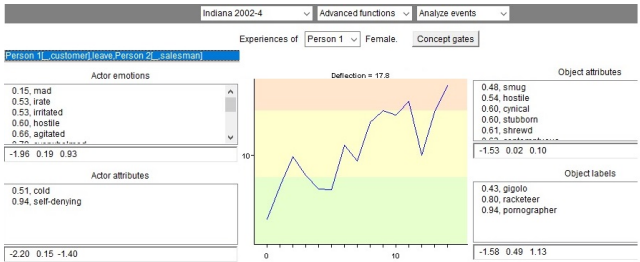

Fig. 2. Interact simulation [3]. This is a screenshot from INTERACT software simulating the interacton between a customer and a salesman.

### C. Identities with modifiers

To minimize deflection in interactions, individuals often generate new behaviors or cognitively redefine an event [7], [11]. When individuals decide to redefine an event, they may most often redefine the behaviors instead of the actors or objects [12].

Given a culture's sentiments (affective dictionary) and impression change equations, ACT allows potential events, emotions, and social labeling to be predicted [13]. For example, an individual will be expected to redefine the features of an ABO event in an attempt to rectify large discrepancies [11]. In the previous example, the nurse forced the aggressive patient to collaborate, has a very different impression.

Adding a modifier to the identities changes the affectiove meaning of the Actor or the object. An aggressive patient has a very different sentiment, and the impression of events for this person is different from a typical patient.

ACT uses amalgamation equations similar to (3) to find the sentiments for the characters with modifiers [14].

$$C_e = -17 + 0.62M_e - 0.14M_p - 0.18M_a + 0.5I_e \quad (3)$$

where, $C_e$, $M_e$, $M_p$, $M_a$, $I_e$ represents evaluation of the modifier-identity, evaluation of the modifier, potency of the modifier, activity of the modifier, and evaluation of the identity respectively. This simple model has a weighted averages of the evaluation for the modifier and the identity. Additionally, potency and activity of the modifier affect the evaluation of the modifier-identity [14].

### D. Word embedding

Calculation of the deflection that underlies predictions about interaction dynamics is dependent on initial cultural sentiments, which are indexed in the affective dictionaries. With the use of word and emoji embedding, we can extend affective dictionaries.

Word embedding is a mapping of words into a numerical vector space that attempts to capture the words' semantic relationship. Word2vec, the most widely used word embedding algorithm, has three million words [15]; a 1000 times more words than commonly used affective dictionaries. Because the proximity of words in the embedded space can imply strong semantic relationships, word embedding has been used as an alternative approach for sentiment surveys [15]–[17]. Building on this, we seek to extend affective dictionaries to include the most commonly used words and emojis in online messaging.

Emojis, which are able to provide affective meaning (e.g., emotion) not contained in words alone, are increasingly being used in online messaging. Emoji2vec developed by Eisner et al. [18] has embeddings for all Unicode emojis. It maps emoji symbols into the same 300-dimensional Word2vec space and can be used together with the Word2vec embedding.

Eisner et al. [18] used the word-embedding of the tags assigned to emojis to derive an emoji-embedding. Table III shows keywords used for some commonly used emojis.

TABLE III
Tags assigned to commonly used emojis [18]

| Index | Emoji | Tags |
|---|---|---|
| 1 | 😡 | rage, irate, grumpy face, mad face, anger, pout, angry face, red face, pouting face |
| 2 | 😎 | smooth, sun, sunny, sunglasses, sunnies, cool, smiling face with sunglasses, mutual best friends, smiling, glasses, snapchat |
| 3 | 😒 | unamused, unamused face, lame, depressed, not amused, unhappy, unimpressed, dissatisfied, disapprove |
| 4 | 😫 | tired face, exhausted, tired, fed up, sleepy |
| 5 | 😂 | laughing, lol, laughing crying, tears, joy, happy, laughing tears, cry, face with tears of joy |
| 6 | 👎 | down, thumb, hand, dislike, gesture, boo, stop, disapproval, sign, thumbs down sign, no, thumbs down |
| 7 | 😍 | heart eyes, smile, flirt, heart face, love, smiling, lovestruck, heart |

## III. Method

Online messaging communication can be viewed from an ACT perspective as comprising the main three characters of Actor, Behavior, and Object. The actor send a message to the object. The action/behavior associated with the message is identified by text mining methods. For example, chatbots use these methods to understand the customer's request and respond accordingly. As we discussed in the previous sections, commonly used affective dictionaries may not include the identified actions. In the first step of emotional tracking, it is necessary to extend the affective dictionaries.

The cost and time constraints of survey instruments have led us to seek alternative approaches to extend the dictionaries. Extended dictionaries will enable researchers to use ACT in many more applications, such as modeling social media behaviors. On the other hand, with the increasing use of emojis in social media communication, not having them in the dictionaries is another limitation.

We used Emoji2vec to extend affective dictionaries to include emojis. Li et al. [17] discussed how regression on word embeddings can outperform other methods in inferring the words' affective meanings. We used regression and a transformation mapping, called translation matrix in this project.

### A. Translation Matrix

Given word embedding $x_i \in \mathbb{R}^{d_1}$ as a source language and the word embedding of its translation in another language, $z_i \in \mathbb{R}^{d_2}$, one intuitive approach for translation is to find a transformation matrix from one embedding to the other. This matrix is known as a translation matrix.

Given the set of vector representation for word pairs, $\{x_i, z_i\}_{i=1}^n$, the goal is to find a transformation matrix $R$ such that $x_i R$ approximates $z_i$. This matrix minimizes the following optimization problem [19],

$$\min_R \sum_{i=1}^{n} ||x_i R - z_i||^2 \qquad (4)$$

This matrix is used to find the translation of new words in the source languages. First, multiply the embedding in the source language to the matrix and then find the closet word in the translation embedding.

### B. Extending the affective dictionaries

We used a translation matrix to map word embedding to an affective space, and then a second-order regression is used to find extended EPA values. We had the following steps to find the extended affective dictionary,

- Word2vec and affective dictionaries represent two different embeddings of words. A set of words from an affective dictionary [20] and their corresponding embedding in word2vec are used as the pair of embedding for machine translation.
- Word embedding is considered as the source language to find the translation matrix.
- 85% of the data is considered as the training set and 15% as the test set.
- The translation matrix in (4) is found from the following equation,

$$x_i R = z_i \to R = (x_i^T x_i)^{-1} (x_i)^T z_i \qquad (5)$$

where $x_i$, $z_i$, $R$ represent the word embedding of the training words in word2vec space, word-embedding of the words in EPA space, and the translation matrix respectively.
- We use a step-wise second order regression to find estimated EPA values from the translated values
- Accuracy of the mapping is calculated on the test set.
- The $R$ matrix and coefficients of the regression model are used to find estimated EPA values for new words in the extended dictionary

We found this approach to provide good accuracy. Fig. 3 compares the correlation between two affective dictionaries [20], [21] with the correlation between our model and the actual EPA values. This shows that our approach provides estimates within the acceptable variability of existing affective dictionaries and provides confidence that new words and emojis have EPA values close to what they would have if additional surveys were performed.

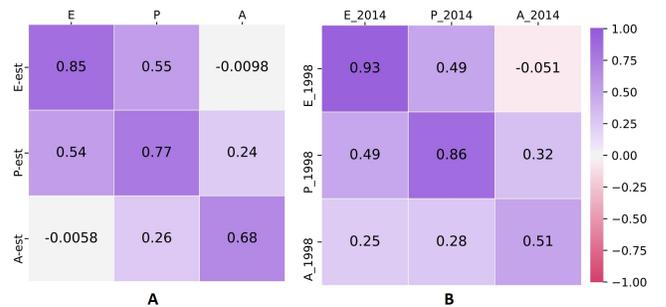

Fig. 3. The correlation between affective dictionaries. A. Estimated values and the affective dictionary collected in 2014 [20]. B. Affective dictionaries collected in 2014 [20] and 1998 [21]

## C. Affective meaning of emojis

We have extended the affective dictionaries using word2vec emedding. To understand emotional feelings during online messaging we need emojis as a part of the dictionary. Since Emoji2vec maps emojis into the same 300-dimensional Word2vec space, the same model gives us an embedding for commonly used emojis. We used the translation matrix found in (5) and the regression parameters to find affective meaning of emojis.

In Table IV, some of these commonly used emojis and estimated values for their affective meaning are given. Note that in affective dictionaries, words are measured on a scale from - 4.3 to + 4.3.

TABLE IV
Estimated EPA values for some commonly used emojis

| Emoji | Evaluation | Potency | Activity | Emoji | Evaluation | Potency | Activity |
|---|---|---|---|---|---|---|---|
|  | 3.65 | 3.21 | 0.55 |  | 0.00 | 1.40 | 1.27 |
|  | 2.45 | 0.24 | -0.65 |  | -0.09 | -0.22 | -0.67 |
|  | 2.35 | 2.24 | 0.50 |  | -0.49 | 1.71 | 0.96 |
|  | 2.06 | 2.01 | 0.64 |  | -0.72 | -0.16 | -0.36 |
|  | 1.93 | 0.58 | -0.32 |  | -0.75 | 1.55 | 2.60 |
|  | 1.86 | 2.16 | 0.60 |  | -0.96 | -1.14 | -1.94 |
|  | 1.80 | 2.50 | 0.58 |  | -1.11 | 0.40 | -1.31 |
|  | 1.40 | 2.33 | 0.40 |  | -1.39 | 0.63 | 1.28 |
|  | 1.33 | 2.74 | 0.87 |  | -1.46 | -0.56 | -0.20 |
|  | 1.33 | 1.26 | 0.64 |  | -1.53 | -1.02 | -1.31 |
|  | 0.48 | 2.56 | -0.00 |  | -1.80 | -0.04 | 0.84 |
|  | 0.36 | 1.68 | 0.02 |  | -2.33 | -1.30 | -0.72 |
|  | 0.14 | 2.05 | 0.93 |  | -2.67 | -1.19 | -0.32 |
|  | 0.05 | 1.08 | -1.45 |  | -2.90 | 0.24 | 1.32 |

Table IV is sorted based on Evaluation dimension. We can intuitively justify this result by comparing pair of emojis. For example, both the nauseated face, 🤢, and the red face, 😡, are negative emojis in the evaluation dimension, but the red face is much stronger and active.

## D. Tracking emotions during a chatbot communication

Given the ability to project an emoji into EPA space, we can model emotional state change during an online messaging communication with the use of emojis. Consider a customer chatting with a chatbot. This scenario is similar to the example shown in Table I but now the customer can use emojis while sending the message.

We used emojis as modifiers for the identity of the customer. Amalgamation equations similar to (3) were used to identify the customer's identity while using emojis to express emotions. We simulate this interaction for four different types of customers: happy, angry, regular, and a customer that uses no emojis. We considered happy, angry and , regular customers add emojis similar to Table V to express their emotions.

The chatbot uses text mining approaches to identify the action tags from the text. Since chatbots understand the action tags for their response, we assumed the action tags column in Table V are given. Using emojis to understand the customer emotion and track it over time is the main focus here.

TABLE V
Interaction of customers and a Chatbot.

| Index | Actor | Chat | Action | Regular | Angry | Happy |
|---|---|---|---|---|---|---|
| 1 | Customer | Hello | greet |  |  |  |
| 2 | Chatbot | Hello, I am the Bot | welcome |  |  |  |
| 3 | Customer | Why my order status is not updated | grouse at |  |  |  |
| 4 | Chatbot | What is your order number | question |  |  |  |
| 5 | Customer | order #8218 | answer |  |  |  |
| 6 | Chatbot | Please wait, I check it | request sth. from |  |  |  |
| 7 | Customer | Take your time | agree with |  |  |  |
| 8 | Chatbot | It will be delivered in two weeks | answer |  |  |  |
| 9 | Customer | Not acceptable, it is too late | criticize |  |  |  |
| 10 | Chatbot | Sorry, I can refund the shipping cost | gratify |  |  |  |
| 11 | Customer | I need full refund | argue with |  |  |  |
| 12 | Chatbot | You get full refund in two business day | uplift |  |  |  |
| 13 | Customer | Sounds good | agree |  |  |  |
| 14 | Chatbot | Thank you for contacting us | thank |  |  |  |
| 15 | Customer | Bye | leaves |  |  |  |

Deflection is defined as the euclidean distance between the sentiments and impressions in EPA space for an ABO interaction. Deflection for the three types of customers is plotted in Fig. 4. We can observe that the deflection increases substantially for all types of customers when the customer grouses, criticizes, or argues with the chatbot. On the other hand, it decreases when the chatbot is offering something to make the customer happy. This trend is similar to what was seen in Fig. 2.

The deflection plot in Fig. 4 shows the deflection for all customer types. To focus on emotional change for each agent, we plot the euclidean distance between sentiments and impressions for the Actor and Object in Fig. 5 which we refer to as the agent's deflection. We can observe that the emotional change for the chatbot while interacting with different customers is very limited; however, the customer's emotional changes are substantial.

Since customer's unsatisfactory was the main source of deflection in Fig. 4, it could be expected that in Fig. 5, the angry customer has the minimum emotional change most of the time. We can go one step further and plot the change of EPA values for all customer types during this interaction.

Fig. 6 shows the EPA values for the customer during the interaction. The rising and falling trends are similar for all customers, but their differences can be observed in each dimension:

- In the evaluation dimension (E), the angry customer always has the minimum evaluation, and the happy customer mostly has the maximum values. The regular customer and the customer that does not use emojis are mostly between the happy and angry customers.
- After initial chit chat, the angry customer is more powerful (P) than the other customers. The other customers are very similar in the potency dimension.
- The activity (A) of the customers does not change that much during this interaction.

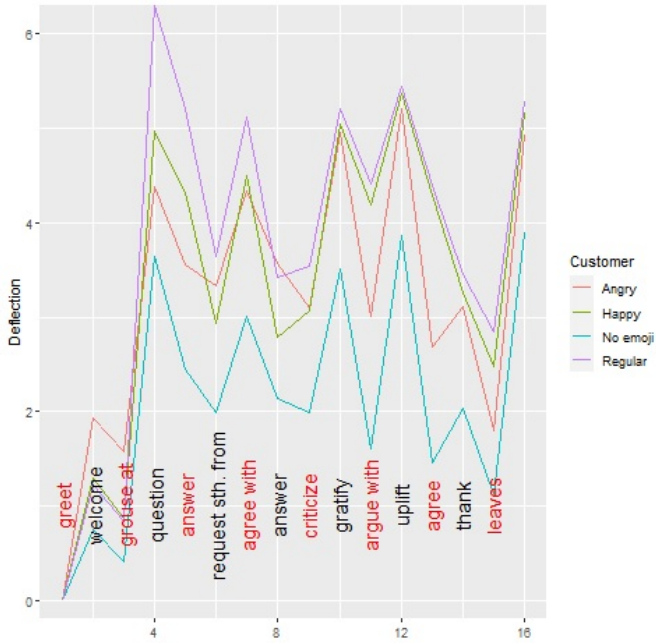

Fig. 4. Deflection for the four types of customers communicating with a chatbot.

The observations from Fig. 6 are similar to what we expect intuitively about the emotional change for these types of customers. We can also see how the impression of chatbot changes in Fig. 7.

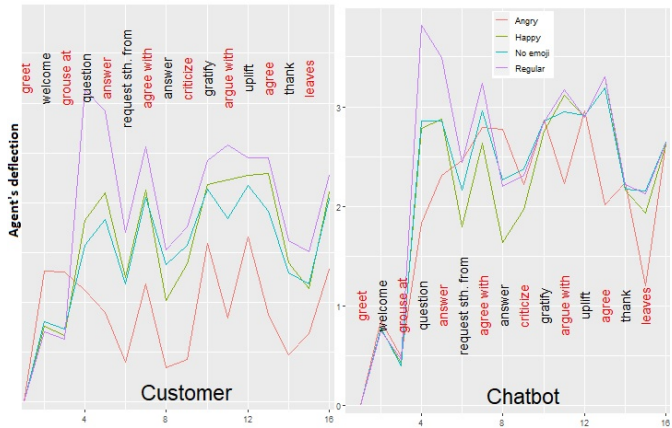

Fig. 5. Deflection for the agent's sentiments

The emotional change for the chatbot in Fig. 7 is less than the customer in Fig. 6. We can observe the following from Fig. 7:

- The evaluation changes substantially when the customer grouses, criticizes, or argues with the chatbot.
- The chatbot is mostly considered to be more powerful when it interacts with an angry customer!
- The impression of chatbot activity does not change substantially for any of the customer types.
- When the chatbot answers a question, it is considered to be more powerful and we see a jump in its potency value.
- We can see when the chatbot questions the customer, its activity will increase significantly.

## IV. Future work

### A. Emoji embedding

As we discussed earlier, Eisner et al. [18] used the word-embedding of the emoji tags to derive an emoji embedding. However, these tags are not necessarily the optimal choices for deriving an affective meaning. For example, Table VI includes tags for two different heart emojis. Some tags such as evil, red, pink, and black may not represent the meanings we expect from a heart emoji.

TABLE VI
Tags assigned to heart emojis

| Index | Emoji | Tags |
|---|---|---|
| 1 | 💗 | growing heart, multiple heart, triple heart |
| 2 | 💕 | love heart, red heart, death, intense, heavy black heart, cold, black, love, pink, romance, passion, heart, evil, desire, red |

Table VII shows the resulting affective meaning of the two heart emojies using the Emoji2vec embedding without removing the tags. As you can see, this is far from expection.

TABLE VII
Estimated EPA values for the heart emojis

| Emoji | Evaluation | Potency | Activity |
|---|---|---|---|
| 💗 | 2.71 | 2.83 | 0.28 |
| 💕 | 0.48 | 2.56 | -0.00 |

In this project, we assumed customers could select emojis from a given list. This assumption is reasonable for the chatbot design. But when it comes to social media analysis, we should a much larger pool of frequently used emojis.

There is limited work on emoji embedding. Creating a survey for finding the affective values for emojis or improving the tags assigned to emojis in Emoji2vec are two possible avenues for future work.

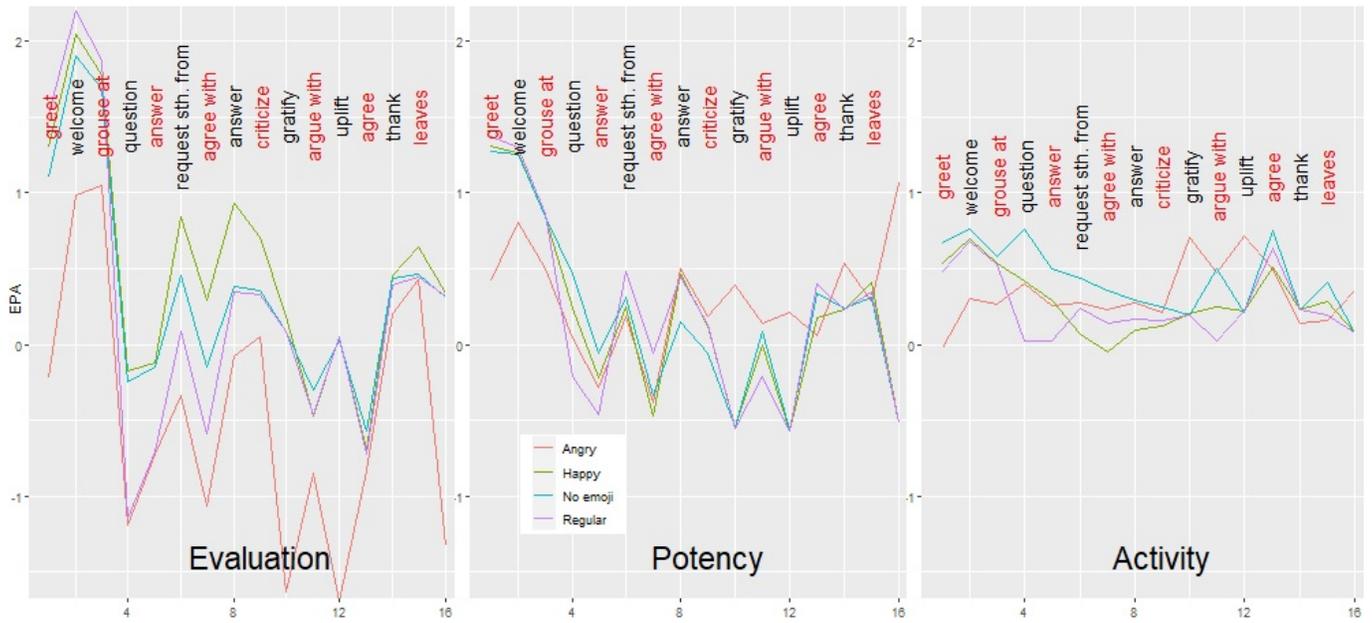

Fig. 6. Estimated EPA values for the customers. As we expect, in the evaluation dimension the happy customer mostly has the top values and the angry one has the lowest values. We can see when the customer complains about the product for all customers the evaluation drops significantly. On the other hand when the chatbot request something from the customer, the customer potency increases significantly

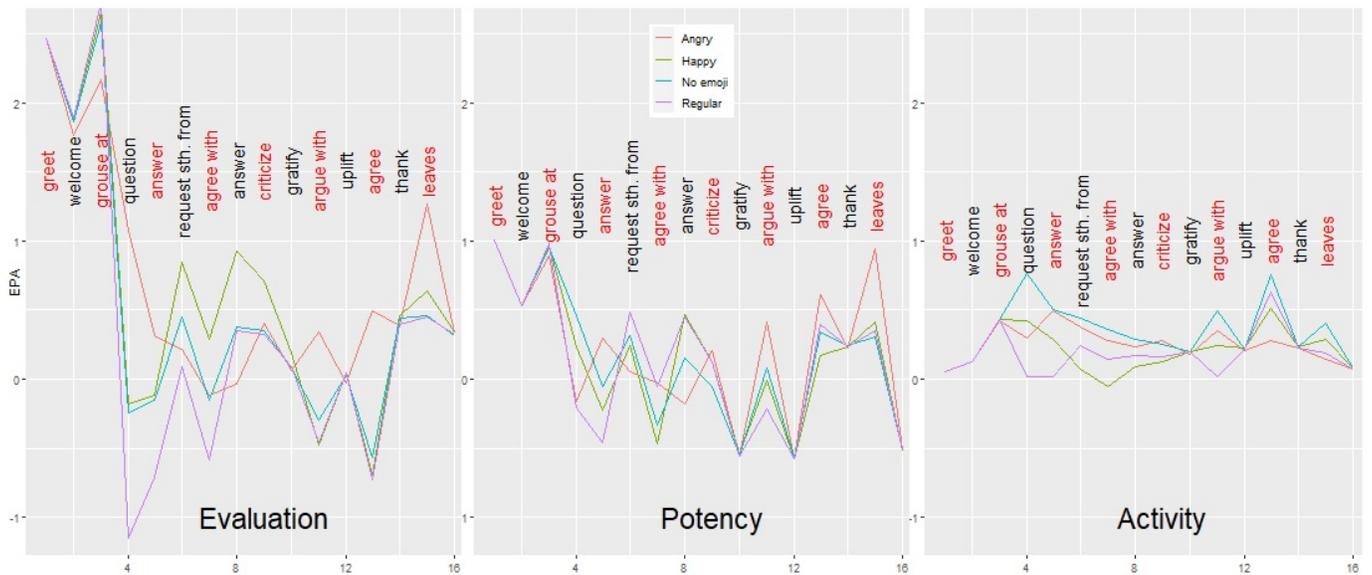

Fig. 7. Estimated EPA values for the chatbot.

### B. Use of emojis for the chatbot

This project assumes the customers could use emojis to express their emotions during the chat. On the other hand, recent research [22] shows that the use of emojis by chatbots can make them more socially attractive to customers. Using the framework introduced here, we can find the optimal emojis for the chatbot to use. Those emojis minimize the deflection in the interaction. This idea is very similar to what INTERACT does to find optimal action, emotion, and attribute during the interaction.

### C. Targets

Having the identity of the Actor and Object, ACT finds the actions that minimize the deflection. In the context of the chatbot, customer satisfaction is an important designing factor. Not only the emojis but also the chatbot's action should make the customer more satisfied. Finding the optimal value function that maximizes customer satisfaction is a subject of future work. Some possible approaches can be considered, such as a weighted summation of deflection terms or changing the target identity to "satisfied customer" instead of the "customer".

## V. Conclusion

To the best of our knowledge, this work is the first to use emoji and word embedding to model emotion change in social media interactions. Also, it is the first time that extended affective dictionaries include emojis.

We have used ACT theories to infer and track the emotional states of individuals during online messaging. The findings from this research can improve chat-bot design and make their communication more naturalistic. Understanding the affective meaning of emojis help machines to have much more efficient conversations with humans.

## Acknowledgment

The authors would like to thank Dawn T. Robinson and Riley King from the University of Georgia and Fateme Nikseresht from the University of Virginia for their great discussions and comments. The authors are also grateful to David R. Heise for allowing the inclusion of INTERACT results in this paper.